\def\comment#1{}

\newcommand{\beg}{\begin{eqnarray}}
\newcommand{\eee}{\end{eqnarray}}

\documentclass[prl,twocolumn,aps]{revtex4}
\usepackage{graphicx}
\def\cm#1{}

\begin{document}
\title{ Andreev-Bashkin effect and  knot solitons in 
interacting mixture of a charged and a neutral superfuids with
 possible releveance for neutron stars}
\author{ Egor Babaev
 \footnote{ http://people.ccmr.cornell.edu/\~{}egor/}
}
\address{Cornell University Clark Hall Ithaca, NY 14853-2501 USA \\
Norwegian University of Science and Technology, Trondheim, N-7491 Norway \\
Institute for Theoretical Physics, Uppsala University  Box 803, S-75108 Uppsala, Sweden 
}
\begin{abstract}
 We discuss  a mixture of  interacting a  neutral and  a charged Bose condensates,
which is supposed being realized in interior of neutron stars
in form of coexistent neutron superfluid and
protonic superconductor. We show
that in this system, besides ordinary vortices of $S^1\rightarrow S^1$
map, the neutron condensate also
 allows for (meta)stable finite-length knotted solitons, which are

characterized by a nontrivial Hopf invariant and in some circumsatnces
are stablized by Faddeev-Skyrme term induced by drag effect. 
We also consider a helical protonic fluxtube 
in this system and show that, in contrast, 
it does not induce a Faddeev-Skyrme term.
\end{abstract}
\maketitle
\newcommand{\la}{\label}
\newcommand{\aaa}{\frac{2 e}{\hbar c}}
\newcommand{\kA}{{\tilde A}}
\newcommand{\bfx}{{\bf \vec x}}
\newcommand{\bfn}{{\bf \vec n}}
\newcommand{\bfE}{{\bf \vec E}}
\newcommand{\bfB}{{\bf \vec B}}
\newcommand{\bfv}{{\bf \vec v}}
\newcommand{\bfU}{{\bf \vec U}}
\newcommand{\ccc}{{\vec{\sf C}}}
\newcommand{\bfp}{{\bf \vec p}}
\newcommand{\f}{\frac}
\newcommand{\bfA}{{\bf \vec A}}
\newcommand{\non}{\nonumber}
\newcommand{\be}{\begin{equation}}
\newcommand{\ee}{\end{equation}}
\newcommand{\ba}{\begin{eqnarray}}
\newcommand{\ea}{\end{eqnarray}}
\newcommand{\bastar}{\begin{eqnarray*}}
\newcommand{\eastar}{\end{eqnarray*}}
\newcommand{\h}{{1 \over 2}}
\section{Introduction}
In a standard model for a neutron star its interior
features superfluidity of
neutron Cooper pairs and superconductivity of proton Cooper
pairs (see e.g. \cite{hoff, peth}).
Both these condensates  allow vortices of $S^1 \rightarrow S^1$ map. 
Earlier it was suggested that  the phenomenon of  glitches
in Crab and Vela pulsars is connected with vortex matter in these stars
\cite{stars}. This remains a topic of intensive studies
and discussions (for recent developments and citations see \cite{link}).
Besides that 
a standard model for a neutron star is 
a special system being a mixture of interacting a charged 
and a neutral  condensates
which makes it  also being a topic of abstract 
academic interest \cite{cl} since such a system
allows for interesting phenomena with no direct counterparts
 in e.g. superconducting metals.
Studies of topological 
defects in a mixture of a charged and a neutral condensates,
 so far, concerned  only ordinary Abrikosov-like
columnar vortices (see e.g. \cite{alpar,sed} and references therein). 
In this paper we argue that, possibly, this is not the 
only one type of stable topological defects allowed in neutron
stars. We show that due to the drag effect in a mixture
of a neutral and a charged superfluid (Andreev-Bashkin effect)
the system also allows under certain conditions stable 
finite-length topological defects characterized by a nontrivial
Hopf invariant, more precisely a special version of
knot solitons.

Finite-length topological defects characterized by a nontrivial Hopf invariant
were attracting interest for a long time in condensed matter physics: earlier
it was discussed in spin-1 neutral superfluids \cite{vol,vol2},
in magnets \cite{cooper}, in  charged
and neutral  two-component Bose condensates  
\cite{we}, \cite{ruo}, \cite{Battye},
in spin-triplet superconductors
\cite{Babaev:2001jt} and in other systems. In neutral systems finite-length 
closed vortices are not stable against shrinkage unless
their size is stabilized by a conservation of some dynamic quantity,
like in  case of a propagating vortex loop.
 A special case is a neutral  two-component system with a phase separation,
where a vortex loop made up of one condensate with confined
 in its core
circulating second condensate is stable
 against shrinkage \cite{Battye}. 
Intrinsically stable topological defects characterized by 
a nontrivial Hopf invariant (the knot solitons) have been discussed in 
the Faddeev nonlinear $O(3)$ sigma model \cite{fadde}
where its stability is ensured by a special fourth-order derivative term:
\be
F_F \ = \ (\partial \bfn)^2 +  \alpha \left(  \bfn \cdot \partial_i
\bfn \times \partial_j\bfn
\right)^2  + \kappa (1-\bfn \cdot \bfn_0)^2,
\la{fadmod}
\ee 
where $\bfn=(n_1,n_2,n_3)$ is a three-component unit vector.
A knot soliton (being in the simplest
case  a toroidal vortex loop)
 is a configuration where the vector $\bfn$ resides
in the core on e.g. the south  pole of the unit sphere, at infinity 
it reaches the  north pole, while in between the core and the 
vortex boundary it performs $n$ rotations if one goes
once around the core and $m$ rotations if one 
goes once along  a closed curve in toroidal
direction.
The stability of knots in this model was extensively 
studied in numerical simulations \cite{nature}. 
Recently it was realized that this model 
is relevant for wide class of physical systems.
First, it was suggested that this model 
may be relevant in the infrared limit of QCD with the knots solitons being 
a  candidate for glueballs \cite{qcd}.
Besides that  an extended version of Faddeev model 
has been derived for two-band superconductors  \cite{we}
and for triplet superconductors \cite{Babaev:2001jt}.

Below we discuss a possibility of formation of finite length 
stable topological defects in a mixture of interacting
charged and neutral Bose condensates.
\section{A mixture of interacting condensates}
A mixture of  a charged (made up of protonic Cooper pairs) and neutral
(made up of neutronic Cooper pairs) Bose condensates
in the interior of neutron stars can be described in the hydrodynamic
limit  by the 
following Ginzburg-Landau functional \cite{alpar,sed}
\beg
F=\f{1}{2}\rho^{pp} {\bf v}_p^2+\f{1}{2}\rho^{nn} {\bf v}_n^2+
\rho^{pn} {\bf v}_p \cdot{\bf v}_n + V + \f{{\bf B}^2}{8\pi}
\la{f}
\eee
where 
${\bf B}$ is magnetic field, and 
\be 
V= a_p|\Psi_p|^2+ \f{b_p}{2}|\Psi_p|^4+ a_n|\Psi_n|^2+ \f{b_n}{2}|\Psi_n|^4+
c|\Psi_p|^2|\Psi_n|^2
\ee
 is the potential term. We begin with a discussion 
of the simplest case of two $s$-wave condensates (so
$\Psi_p=|\Psi_p|e^{i \phi_p}$
and $\Psi_n=|\Psi_n|e^{i \phi_n}$ are complex scalar fields
which discribe proton and neutron condensates correspondingly).
 In the above expression 
\be 
{\bf v}_n=(\hbar/2m_n )\nabla \phi_n
\ee
 and 
\be
{\bf v}_p=(\hbar/2m_p) \nabla \phi_p-(2e/m_pc) {\bf A}
\ee
are superfluid velocities of neutron and proton condensates.
The key feature of this system is the Andreev-Bashkin effect
\cite{andr}: due to interaction between two superfluids the particle
current of one of the condensates is carried by the superfluid velocity of another
so the superfluid mass current of protons  and neutrons in such a system is \cite{andr,alpar,sed}:
\beg
{\bf w}_p= \rho^{pp}{\bf v}_p +  \rho^{pn}{\bf v}_n; \nonumber \\
{\bf w}_n= \rho^{nn}{\bf v}_n +  \rho^{np}{\bf v}_p,
\la{gg}
\eee
where $\rho^{pn}=\rho^{np}$ is the superfluid density of one of the 
condensates which is carried by superfluid velocity of another.
Because of the Andreev-Bashkin effect  the  charged
supercurrent  in this system depends on gradients
of neutron condensate (as it follows from (\ref{f}),(\ref{gg})):
\be
{\bf J}=
\f{e\hbar\rho^{pp} }{m_p^2}\left(\f{\rho^{pn}m_p}{\rho^{pp} m_n }\nabla \phi_n+
 \nabla \phi_p    - \f{4e}{c\hbar} {\bf A}
\right)
\la{cur}
\ee
Let us discuss topological defects, allowed in (\ref{f}), other than Abrikosov vortices.
\section{Helical neutron vortex loop}
Let us consider a vortex loop made up of neutron condensate
with  zero density of neutron Cooper pairs in its core. Let us 
introduce a new variable $\theta$ as follows:
$\f{\rho^{pn}m_p}{\rho^{pp} m_n }=\sin^2\big(\f{\theta}{2}\big)$.
We will consider a defect where 
 if we go from the core center to the boundary of the fluxtube
in a cross section to the vortex, the variable
$\theta$ grows from $0$ to $\pi$.
Since at the center of the vortex we have chosen that 
the density of the neutron condensate vanishes
then indeed there is no drag effect in the center of the fluxtube and
correspondingly $\rho^{pn}$ is zero in the core.
This allows one to 
chose the boundary condition $\sin^2(\theta/2)=0$ in the 
center of the vortex.
Let us now impose
the following  configuration of $\phi_n$: if we go once around the 
vortex core the  $\phi_n$ changes $2\pi n$,
while if we cover the vortex loop once in toroidal
direction (a closed curve along the core)
$\phi_n$ changes $2 \pi l$
with $n,l$ being integer. 
Such a situation naturally occurs if a loop is formed
around rotation-induced  vortex line 
or in case of two interlinked loops.
This configuration
corresponds to a spiral superflow of the neutron Cooper
pairs in such a vortex ring. Topologically such a vortex
is equivalent to knot solitons considered in \cite{we} and
 can also be characterized by a unit vector
 $\vec{\bf  e} =(\cos\phi_n\sin\theta,\sin\phi_n\sin\theta,\cos\theta)$
with a nontrivial winding.
We stress that we do not impose a nontrivial
winding on $\phi_p$
(compare with discussion of knot solitons in the two-gap
model \cite{we} where, in contrast, in a knot soliton the  phases of both condensates
must  have a nontrivial winding number, however, as discussed below, neutral-charged
mixture is principally different from the system in Ref. \cite{we}).

Indeed the nontrivial superflow of neutron Cooper pairs
induces a drag current of proton Cooper pairs
which in turn induces a magnetic field
which can be calculated from (\ref{cur}):
\beg
&&{\bf B}=
{\rm curl} \left[-{\bf J} \f{c m_p^2}{4 e^2 \rho^{pp}}+\f{c\hbar}{4e}\f{\rho^{pn}m_p}{\rho^{pp} m_n }\nabla \phi_n
\right] ,
\eee
which can also be written as
\beg 
 B_k& =&
-\f{c m_p^2}{4 e^2 \rho^{pp}} [\nabla_i J_j - \nabla_j J_i]
\nonumber \\ 
&& +\f{c\hbar}{8e}\sin\theta [\nabla_i \theta \nabla_j \phi_n-\nabla_j \theta \nabla_i \phi_n ]
\la{b}
\eee
This self-induced magnetic field 
gives the following contribution to the free energy (\ref{f}):
\beg
F_m&=& \f{{\bf B}^2}{8\pi}=
\f{c^2\hbar^2}{512\pi e^2}\Bigg[\f{ 2 m_p^2}{\hbar^2 e \rho^{pp}} [\nabla_i J_j - \nabla_j J_i]
\nonumber \\
&-&\sin\theta [\nabla_i \theta \nabla_j \phi_n-\nabla_j \theta \nabla_i \phi_n ]
\Bigg]^2
\la{fm}
\eee
Which is a version of the Faddeev fourth-order derivative term analogous to the fourth-order
derivative term in \cite{we,Babaev:2001jt} closely related to the
fourth-order derivative term in (\ref{fadmod}). 
The fourth order derivative terms of this type  provide  stability to finite length topological
defects \cite{fadde,nature}.
Physically, in a mixture of a charged and a neutral condensates
this effect  corresponds
to the following situation: as mentioned above, the nontrivial configuration of phase and 
density of neutron condensate induces a charged drag current of 
proton Cooper pairs which results in the configuration of magnetic field
(\ref{b}). This configuration has the special  feature that if the vortex shrinks then
the magnitude of the self-induced magnetic field grows. 
We also remark that $\rho^{pp}$ is  a measure
of background density of proton condensate
which is not required to vary to produce a knot soliton.

In the two-gap model in \cite{we} there is a competition of 
the fourth-order derivative term (which corresponds to self-induced
magnetic field) versus  a second-order derivative term and 
 a mass term for the third component 
of the $O(3)$-symmetric order parameter $\bfn$ 
(the third component of $\bfn$ is related to condensate densities
in \cite{we}
and thus it is massive).
In contrast, in the present model in the competition 
also participates  kinetic energy
of superflow of neutron Cooper pairs (which is minimized  if 
the vortex shrinks). 
  A  necessary condition for (meta)stability of such a vortex loop
is that the competition of 
kinetic energy of  superflows, gradients of condensate density
versus the self-induced magnetic field 
would stabilize the vortex loop at a length scale which 
corresponds to magnitude of magnetic field $|{\bf B}({\bf x})|$ 
smaller than the field which could break proton Cooper pairs.
We also emphasis that one of the differences with the 
system of two charged scalar fields in \cite{we} is that 
in the present case the self induced magnetic field
comes from drag current in the vicinity of the core while
the superflow of neutron Cooper pairs is extended (not  localized
on length scale shorter or equal to penetration length like
the field inducing currents in \cite{we}).
We also remark that indeed  the effective action (\ref{f})
is assumed being derived from a microscopic theory in the approximation 
of small gradients. Indeed one can derive higher-order
derivative terms from a microscopic theory but this 
sort of  terms, in contrast to the term 
(\ref{fm}) is irrelevant for
discussion of the stability of finite-length 
topological defects in this system. Indeed  a competition 
between second- and fourth-order derivative terms 
obtained in a derivative expansion would stabilize
a topological defect at a characteristic  length scale where all 
the higher-order derivative terms become of the same order
of magnitude. So, at such length scales the derivative 
expansion fails.
We also would like to stress that 
in the present system the knot soliton is prevented 
against a collapse by a finite energy barrier, in 
contrast to an infinite energy barrier
in the case of the Faddeev's nonlinear $\sigma$-model
considered in mathematical physics \cite{fadde}.
That is, a zero in proton condensate density,
outside core, may lead to unwinding of a knot soliton
since in a such a point the unit vector $\bfn$ is ill-defined and
thus the Hopf map is ill-defined as well. However the modulus
of proton condensate order parameter is massive so producing
such a singularity is energetically expensive. Thus, it is a 
finite energy barrier which prevents a knot soliton
in a superfluid/superconductor from collapsing.
\section{An example of generalization to other pairing symmetries}
Let us generalize the discussion to the case of 
a mixture of a spin-triplet neutron condensate and 
a singlet proton condensate in order to show that 
the picture does not depend significantly
on pairing symmetry.
The order parameter of the spin-1 neutral condensate is
$ |\Psi_n ({\bf x})|^2 \zeta_q ({\bf x})$
where  $(q=1,0,-1)$ and $\zeta$ is
a normalized spinor $\zeta^\dagger \cdot \zeta=1$.
Free energy of a  neutral spin-1 system is (see e.g. \cite{ho}):
\beg
F_t&=& \f{\hbar^2}{2 m_n} (\nabla |\Psi_n| )^2 +
 \f{\hbar^2}{2m_n}|\Psi_n|^2 (\nabla \zeta)^2 - \mu |\Psi_n|^2 \nonumber \\&+&\f{|\Psi_n|^4}{2} \left[ c_0 +c_2 <{\bf S}>^2\right], 
\label{neu}
\eee
where  $<{\bf S}> =\zeta_q^*{\bf S}_{q j}\zeta_j$ is spin.
Degenerate spinors  are related to
each other by gauge transformation $e^{i\phi_n}$ and spin rotations
${\cal U}(\alpha, \beta, \tau)$$=$$e^{-iF_{z}\alpha} e^{-iF_{y}\beta}
e^{-iF_{z}\tau}$,  where $(\alpha, \beta, \tau)$ are the
Euler angles. The topological defects in the neutral
system  like this have been intensively studied (see e.g. \cite{vol,vol2}).
A charged counterpart of this system in ferromagnetic state allows stable knot 
solitons as it was shown in \cite{Babaev:2001jt}.

Let us consider first the   ferromagnetic state (which  emerges when
$c_{2}<0$) in context of a mixture
of superfluids. The energy in this case is minimized by 
$<{\bf S}>^2=1$ and the ground state spinor and density are \cite{ho}:
$\zeta =
 e^{i(\phi_n-\tau)} ( e^{-i\alpha}{\rm cos}^{2}\frac{\beta}{2}, 
\sqrt{2} {\rm cos}\frac{\beta}{2}{\rm sin}\frac{\beta}{2},
 e^{i\alpha}{\rm sin}^{2}\frac{\beta}{2} ); \ \
|\Psi_n|^2 = \frac{1}{c_0+c_{2}}\mu$.
The superfluid velocity in ferromagnetic case  is \cite{ho}: 
${\bf v}_n = 
 \f{\hbar}{2m_n}[\nabla (\phi_n-\tau)-\cos\beta \nabla \alpha]$. 
So in a mixture of a neutral  ferromagnetic triplet condensate 
and a charged singlet condensate  the equation for charged  current
is:
\beg
{\bf J}&=&
\f{e\hbar}{m_p}\f{\rho^{pp} }{m_p}
\Big(\f{\rho^{pn}m_p}{\rho^{pp} m_n }[\nabla (\phi_n-\tau)-\cos\beta \nabla \alpha]
\nonumber \\
&+& \nabla \phi_p    - \f{4e}{c\hbar} {\bf A}
\Big)
\la{curt}
\eee
From this expression we can see that assuming e.g. that
there is no nontrivial windings in the variables $\alpha$ and $\beta$,
the system  reduces to (\ref{cur}) and thus allows for
the topological defects in the form
described in the first part of the paper. We emphasis that 
there are no knots of this type in a  charged
ferromagnetic triplet system considered in  \cite{Babaev:2001jt}
because 
in the current equation of a  charged triplet superconductor,
the ratio of the coefficients in front of the  vector
potential term and the gradient  term analogous to $\nabla (\phi_n-\tau)$
does not depend on the carrier density and thus one can not obtain 
a contribution analogous to Faddeev term to the  free energy 
by imposing a nontrivial configuration of 
the first gradient term in the current equation 
similar to (\ref{curt}) in the system \cite{Babaev:2001jt}.
 In a charged triplet case the knot 
soliton may form only as a  spin texture \cite{Babaev:2001jt}.
So a neutral-charged mixture with drag effect in its magnetic
properties is principally different from a genuine charged system.
Spin-texture knots can be formed
in the present system too, as a configuration of the order parameter
$\vec{\bf s} = (\cos\beta\sin\alpha,\sin\beta\sin\alpha,\cos\beta)$
characterized by a nontrivial Hopf invariant. Such a texture
generates  magnetic field due  drag current induced
by the  superflow of 
the neutron Cooper pairs, which is produced by the spin texture. 
So, in general, there is  the following
nontrivial  magnetic energy contribution to the free energy 
functional:
\beg
F_m^t&=& 
\f{c^2\hbar^2}{512\pi e^2}\Bigg[\f{2 m_p^2}{\hbar^2 e \rho^{pp}} [\nabla_i J_j - \nabla_j J_i]
-\sin\theta [\nabla_i \theta \nabla_j \phi_n
\nonumber \\&&
-\nabla_j \theta \nabla_i \phi_n ]
- \sin\beta [\nabla_i \beta \nabla_j \alpha-\nabla_j \beta \nabla_i 
\alpha ] 
\Bigg]^2
\la{fmt}
\eee
It must be observed that the spin-texture knot 
soliton is structurally   different from the
topologically equivalent knot of the type considered in the 
first part of the paper. The spin-texture knot is coreless (there are no 
zeroes of the condensate density in it).
The third component of the order parameter
$\vec{\bf s} = (\cos\beta\sin\alpha,\sin\beta\sin\alpha,\cos\beta)$
is massless in this case, thus the spin-texture knot solitons in this
system are energetically less expensive 
and have larger characteristic size than the topologically
equivalent knots in the variable
$\vec{\bf  e} =(\cos\phi_n\sin\theta,\sin\phi_n\sin\theta,\cos\theta)$.

Let us now  consider the ``polar" phase 
of  triplet superconductors which
 is the case when $c_{2}>0$ in (\ref{neu}).
The energy is minimized then by $<{\bf S}>=0$. The 
spinor $\zeta$ and the condensate density in the ground state are \cite{ho}:
$\zeta=e^{i\phi_n} ( -\frac{1}{\sqrt{2}}e^{-i\alpha}{\rm sin}\beta, 
{\rm cos}\beta, \frac{1}{\sqrt{2}}e^{i\alpha}
{\rm sin}\beta  ); \ {|\Psi_n|^2}= \mu/{c_{o}}$.
The  superfluid velocity in this case  is (see e.g. \cite{ho}):
$ {\bf v}_n =  \f{\hbar }{2 m_n}\nabla \phi_n$ which
is analogous to singlet case. Thus in  the antiferromagnetic 
case the allowed knot solitons are equivalent to knots in a mixture
of two singlet condensates considered in the first 
part of the paper.

We also remark that 
it is generally assumed that there is no proton-neutron pairing in a neutron star
becasue of  large differences
in their Fermi energies. 

While we can not make at this stage any definite predicatoins 
(which would require large-scale numerical simulations),
let us however discuss possible mechanisms of 
formation of knot solitons of the discussed above
types  in neutron stars.
As it is known, ordinary vortices in superconductors
form e.g. as an energetically preferred state in external magnetic field.
Indeed it is not the only possible mechanism of 
creation of topological defects. For instance many defects 
in liquid helium are created in a laboratory without rotation,
by various thermal quench techniques including neutron irradiation
since a symmetry breaking phase transition is accompanied
by creation of topological defects. In case of ordinary 
Abelian Higgs model, created during transition vortex loops shrink,
while the knot solitons should remain stable.
Another less common
mechanism of formation of vortices is  the spontaneous
vortex state \cite{spvor} which emerges when in a system coexist
superconductivity and magnetism and since a vortex 
carries magnetic field it may have negative 
contribution to 
free energy functional via Zeeman-like
 coupling terms. So
 for such systems it is energetically
preferred to form  vortex lattice 
even without applied external fields \cite{spvor}.
Similar mechanism may work in neutron stars if 
in a presence of a spin-1 superfluid, there is a direct coupling 
of spins of Cooper pairs to magnetic field. In general
an effective action of triplet superfluid
features a Zeeman-like term
which is a direct coupling of spin $\bf S$
to magnetic field $\bf B$
\be
F_Z= - \eta {\bf S } \cdot { \bf B} 
\ee
Indeed existence of such a term could 
result in a range of parameters where 
knots would have a finite negative energy 
if spins of neutron Cooper pairs
in the knot soliton are aligned along the self-induced
magnetic field. A definite answer to this
question is however a complicated problem  
%
because of competition of many terms.
This
appears being a particularly  interesting problem
for numerical simulations. Since in a knot, the magnetic
field grows if the knot shrinks, it could 
be that in a such a system a formation 
of a dense ensemble of knot solitons 
is energetically preferred over a spontaneous
vortex state of Abrikosov vortices.
A definite answer to this question may
however be only obtained in a large scale 
numerical simulation.
Thus, if an ensemble of knot
solitions is formed in a neutron star then
one of the apparent consequences 
would be its interaction with 
ordinary columnar neutron vortices, then apparently 
in such a case
knot solitons would disturb  a regular
lattice of neutron vortices.
\section{Helical protonic fluxtube formed around a neutron vortex}
Above we considered knot solitons which appear due to notrivial helical
 windings of neutron condensate phase. 
In principle there is a theoretical mechamism 
which would allow formation of helical 
vortex loops of proton condensate. Let us 
show however that  a helical protonic vortex loop is not  a
knot soliton and it is not stable.
Here we stress 
the most recent studies \cite{link}
indicate type-I behaviour of proton condensates.
Let us now however consider the model \cite{ruderman}. In that model
a neutron star possesses a lattice of uniform neutron vortices
and a complicated structure of sparse entangled 
proton flux tubes (see Fig. 3 in \cite{ruderman}).
In the dymanical processes discussed in \cite{ruderman}
one should expect  that entangled proton fluxtubes 
may dymanically form  rings around columnar neutron vortices
as shown on Fig 1.
\begin{figure} 
\includegraphics[width=0.5\columnwidth]{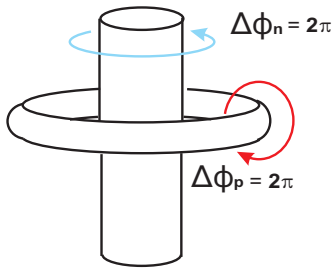}
\caption{A protonic fluxtube ring around neutron vortex. Due to drag
by neutron Cooper pairs, the resulting charge current in protonic fluxtube is
helical.}
\label{fig1}
\end{figure}
 Let us  now 
consider such a ring. 
The charge current in such a ring is given
by eq. (\ref{cur}). When we go around
such a flux tube, the protonic phase
$\phi_p$ changes $2 \pi$, however 
there is also a current along such a vortex
due to drag effect by superfluid neutron Cooper
pairs which is characterized by a nontrivial
phase winding of $\phi_n$ which 
changes $2\pi$ when we cover flux tube
once in toroidal direction. This results in 
a spiral net charge current in such a vortex loop
resembling that of a knot solition
considered in the first part of the paper.
Let us show however that such a  vortex 
loop is not stable:
The magnetic field in such a helical 
fluxtube is given by
\be
{\bf B} = \f{c\hbar}{4e} {\rm curl}
\left[ -{\bf J}\f{m_p^2}{e\hbar\rho^{pp} } - \nabla \phi_p -
\f{\rho^{pn}m_p}{\rho^{pp} m_n }\nabla \phi_n \right]
\ee
In such a configuration, in spite of helical 
net charge current, the individual phase 
configurations of $\phi_p$ and $\phi_n$
are not helical, besides that, the ratio of the vector
potential term to gradient term for
$\phi_p$ is constant. Thus such a helical superflow
does not result into a self-induced Faddeev-Skyrme-like
term, which, if it would be present, would 
significanly affect the considerations in \cite{ruderman,link}.
\section{Conclusion}

In conclusion we studied topological defects other
than Abrikosov vortices in an interacting mixture of a 
neutral and a charged condensates. Such a system 
is believed being realized in the interior of neutron stars.  
We have shown that due to Andreev-Bashkin effect 
the system possesses a large variety of  knot  solitons
of different nature than  the  knot solitons in the
systems studied before. We also suggested that 
due to Zeeman coupling term,  there could be a theoretical 
possibility of an exotic inhomogeneous ground state
in this system: a spontaneous formation 
of a  dense ensemble of knot solitons. 

It is a great pleasure to thank   L.D. Faddeev, A. J. Niemi,
G.E. Volovik,  B. Carter,   
C. Pethick for many useful discussions. 
We also acknowledge support by Goran Gustafsson Stiftelse UU/KTH.
This work was also partially supported 
by the Research Council of Norway, Grant No.
157798/432,  STINT and  the Swedish Research Council and 
National Science Foundation Grant DMR-0302347.

 
\end{document}